ONE METHOD FOR BUILDING NDNLOCAL DENSITY FUNCTIONALS
FOR ENERGY

I. I. Mazin

A new method is proposed for constructing energy density functionals which include a nonlocal dependence on the density gradients. This method is used to construct functionals for kinetic energy which is a nonlocal generalization of the Thomas-Fermi-Kirzhnits functional.

The density functional theory (DFT) [1] is a powerful means for quantitative investigation of the properties of atoms, molecules, and solids. In the DFT, the total energy of the ground state of a system of electrons in the external potential $V_{ext}$ is described by the formula

$$E[\rho] = T_S[\rho] + E_H[\rho] + \int V_{ext}(\mathbf{r})\rho(\mathbf{r})d\mathbf{r} + E_{xc}[\rho], \qquad (1)$$

where $\rho(\mathbf{r})$ is the electron density; $E_H$ is the Hartree energy; $T_s$ is the kinetic energy of interacting fermions with the same density $\rho(\mathbf{r})$; and $E_{xc}$ is the exchange-correlation energy. The DFT is normally used in a Kohn-Sham formulation [1], in which $T_S[\rho]$ is calculated precisely, through solution of the effective single-electron Schroedinger equation for noninteracting fermions. This is associated with the absence of sufficiently precise explicit functionals for $T_S[\rho]$.

The following are the known expressions for $T_S[\rho]$: 1. The Thomas-Fermi functional $T_{TF}[\rho] = 0.3(3\pi)^{\frac{2}{3}} \int \rho^{\frac{5}{3}} d\mathbf{r}$ 2. The Weizsacker-Kirzhnits functional $T_W[\rho] = (\lambda/8) \int (\nabla\rho/\rho)^2 \rho d\mathbf{r}$, $T_{TF} + T_W$. The coefficient $\lambda$ in Weizsacker's original work is equal to one. As Kirzhnits showed in [2], when $T_W$ is understood as the first term in the gradient expansion of $T_s$, then $\lambda = 1/9$. 3. Nonlocal functionals.

The author examines these functionals from the standpoint of the linear response theory. The following relation is known:

$$\frac{\delta^2 E[\rho]}{\delta\rho(\mathbf{r})\delta\rho(\mathbf{r}')} = -\chi^{-1}(\mathbf{r},\mathbf{r}'),$$

$\chi^{-1}$ is the inverse electron susceptibility. In particular, with an almost Uniform electron gas [1]

$$\frac{\delta^2 T_s[\rho]}{\delta\rho(\mathbf{r})\delta\rho(\mathbf{r}')} = -\chi_0^{-1}(\mathbf{r},\mathbf{r}'),$$

$$\chi^{-1}(\mathbf{r},\mathbf{r}') = \chi_0^{-1}(\mathbf{r},\mathbf{r}') - V_o(\mathbf{r}-\mathbf{r}') - \delta^2 E_{xc}/\delta\rho(\mathbf{r})\delta\rho(\mathbf{r}'), \qquad (3)$$

where $\chi_0$ is the normal "Lindhard" susceptibility of noninteracting electrons and $V_0$ is the Coulomb potential. Since $\chi_0$ and $\chi$ for a uniform gas are well known, the requirement (3) places substantial limitations on the type of the functionals $T_S$ and $E_{xc}$. While there 'arefunctionals for Exc which satisfy (3) with reasonable precision [1], for $T_S$ there are essentially none. In fact, in the momentum space

$$\chi_{TF}^{-1}(k) + \chi_W^{-1}(k) = \pi^2/k_F + 6\pi^2\lambda k^2/k_F^3 \qquad (4)$$

where $k_F = (3\pi^2)^{1/3}$. At the same time, according to (3), this value must be equal to $\chi_0^{-1}(k)$, i.e.,

$$\chi_0^{-1}(k) = \frac{2\pi^2}{k_F}[1 + \left(\frac{1-\eta^2}{2\eta}\right)\ln\left|\frac{1+\eta}{1-\eta}\right|]^{-1}. \qquad (5)$$

In a range of $k \to 0$, (4) coincides with (5) for λ=1/9, while λ=1 is the case in a range of $k \to \infty$. It is apparent that the Kirzhnitz-Weizsacker functional cannot correctly describe an almost uniform electron gas. There is an attempt in the literature [3] to construct a nonlocal functional of the following type:

$$T_s[\rho] = T_{TF}[\tilde{\rho}] + T_W[\tilde{\rho}], \quad \tilde{\rho} = \int w(\mathbf{r}-\mathbf{r}')\rho(\mathbf{r}')d\mathbf{r}'.$$

where the function $w$ is selected from condition (3). In the author's opinion, this method is not physically natural. It is desirable to preserve the appearance similar to (4) with consideration, however, of the relation $\lambda(k)$, according to (5). It is easy to see that selection

$$T_W = \frac{1}{8}\int \frac{\nabla\rho(\mathbf{r})}{\rho(\mathbf{r})}\cdot\frac{\nabla\rho(\mathbf{r}')}{\rho(\mathbf{r}')}\lambda(\mathbf{r}-\mathbf{r}',\bar{\rho})\,d\mathbf{r}d\mathbf{r}' \qquad (6)$$

where $\bar{\rho}$ is the function of **r**, **r**' such that in a range of $\rho(\mathbf{r}) \to$ const, $\bar{\rho} \to \rho$, and $\lambda(\mathbf{r}-\mathbf{r}')$ is the Fourier transform of $\lambda(k)$ which ensures fulfillment of (3). Selection

$$\lambda(r) \approx \delta(r) - [2k_F^2(\bar{\rho})C/9\pi r]\exp[-2k_F(\bar{\rho})\sqrt{C}r]. \qquad (8)$$

where $C = 1/40$ ensures good precision of fulfillment of (3). It may be expected that by preserving the general structure of the first gradient correction and being precise within an almost uniform density, formula (8) will be sufficiently precise for actual systems as well.

In [4], Herring proposed the use of a one-dimensional model for testing the density functionals, since the properties of the functional are normally not chiefly dependent on the dimensionality of the space. In the one-dimensional case the functional for $T_S$ which corresponds to (7), (8) appears as follows:

Ratio of the Kinetic Energy Calculated Using Different
Functionals to the Precise Energy for Different
Potentials and the Numbers of Electrons

| Potential type | Number of particles | $\lambda_0$ – Thomas-Fermi functional | $\lambda_W$ – Weizsacker gradient correction | $\lambda_K$ – Kirzhnits gradient correction | Nonlocal functional |
|---|---|---|---|---|---|
| Box | 2 | 0.833 | 1.833 | 0.500 | 1.197 |
|  | 10 | 0.939 | 1.200 | 0.852 | 1.028 |
|  | 20 | 0.966 | 1.103 | 0.921 | 1.012 |
|  | 30 | 0.967 | 1.069 | 0.946 | 1.008 |
| Harmonic potential | 2 | 1.209 | 2.209 | 0.876 | 1.287 |
|  | 10 | 1.018 | 1.100 | 0.991 | 1.014 |
|  | 20 | 1.006 | 1.033 | 0.997 | 1.004 |
|  | 30 | 1.003 | 1.014 | 0.999 | 1.002 |
| $\delta$ | 2 | 1.089 | 2.089 | 0.756 | 1.108 |

$$T_{TF}[\rho] = \left(\frac{\pi}{2}\right)^3 \frac{1}{3\pi} \int \rho^3 \, d\mathbf{r}, \qquad \lambda(r) = \delta(r) - \frac{8\pi}{3} \exp(-2\pi\bar{\rho}r). \qquad (9)$$

The maximal values of $\lambda$ here are $\lambda_W = 1$ and $\lambda_K = -1/3$. As in the three-dimensional case, here it is convenient to use $\bar{\rho}\sqrt{\rho(\mathbf{r})\rho(\mathbf{r}')}$ as $\bar{\rho}$; (9) was tested in a system of electrons in a potential box, in a harmonic potential, and for two electrons in a o-shaped well. The results are presented in Table 1. It is evident that the natural approximation which ensures sufficiently high precision for all three types of potential is the nonlocal functional (9). It is noted that an analogous method may be used for constructing nonlocal functionals for exchange-correlation energy as well.

17 September 1987